\documentclass[a4paper,10pt]{article}
\newcommand{\m}{\mu}
\newcommand{\n}{\nu}

\def\equ#1{(\ref{#1})}
\begin{document}

\title{Fuzzy spaces topology change as a possible solution to the black hole information loss paradox}
\author{C.A.S. Silva \thanks{calex@fisica.ufc.br}} 

\maketitle

\begin{abstract}
In this paper we argue that modelling the black hole event horizon as a fuzzy sphere we can get a possible solution to the black hole information loss paradox.

\end{abstract}

\section{Introduction}

\indent One question have haunted the mind of physicists in the recent years: what is the ultimate fate of something that falls into a black hole? Is it crushed out of existence at a singularity, or does it end up ``somewere else''? From these questions arises the black hole infomation loss paradox \cite{sw.hawking-cmp43, sw.hawking-prd14, j.preskill-hepth9209058,dn.page-hepth9305040}.

Bekenstein and Hawking have been shown that the surface of a black hole should be quantized in multiple of the Planck area \cite{jd.bekenstein-lnc4}. Quantizing the event horizon is related with the idea of fuzzy sphere, in which points are ``smeared out'' and the geometry becomes non-local \cite{j.madore-aincdgpa}. It was suggested some time ago that a black hole event horizon might be modeled by a fuzzy sphere \cite{j.madore-aincdgpa, bp.dolan-jhep0502}, in a way that one can ask if fuzzy sphere model could help us to solve the black hole infomation loss paradox cited above. 

Fuzzy sphere are compact noncommutative manifolds which has been treated extensively in the literature due the natural realization of the spacetime uncertainty relation by its noncommutative geometry and its connection with M-theory. Besides one can use fuzzy spheres as a regularization scheme alternative to the lattice regularization \cite{h.grosse-ijtp35}. Unlike the lattice, fuzzy spheres preserve the continuous symmetries of the space-time considered, and hence it is expected that the situation concerning chiral symmetry and supersymmetry might be ameliorated \cite{h.grosse-plb283,h.grosse-lmp33,h.grosse-cmp185}.

Balachandran et al \cite{ap.balachandran-ijmpa19} have shown that fuzzy spaces posses a inherent Hopf algebra, and a topology change process where a fuzzy sphere splits in  two others can be defined. In this paper we shall try to resolve the information loss paradox using this process. There a new topologically disconnected region appear where information can be stored.

Topology change has been used as a proposal to solve the black hole information paradox, but it has found out some obstacles. In the reference \cite{sdh.hsu-plb644} the main objections against topology change process are addressed.
In this paper we argue that using the fuzzy sphere model the objections against topology change can be solved in a precise way, and then we shall get to find a possible solution to the black hole information loss paradox.

\section{Information loss paradox}
Hawking has been showing that black holes emit radiation. This establish a deep and satisfying connection between gravitation, quantum theory, and thermodynamics. In this framework black holes have an entropy which is given by:
\begin{equation}
S = \frac{k_{b}A}{4l^{2}_{p}} \label{B-H-form}.
\end{equation}

We have that, in statistical physics, entropy counts the number of accessible microstates that a system can occupy, where all states are presumed to occur with equal probability. On the other hand, black holes can be completely characterised by only three externally observable classical parameters: mass, electric charge, and angular momentum. All other information about the matter which formed a black hole or is falling into it, "disappears" behind the black-hole event horizon in a way that the nature of these microstates is obscure. We must to think if a observer outside of the black hole could have, at any time, some information about the initial state of the matter which collapsed into the black hole. We have that, as the black hole emits radiation, it loses mass till to evaporates completely and, at finish, the only thing we have is the Hawking radiation. Then the answer to the question above depends on the nature of this radiation. 

Hawking firstly showed that the black hole radiation is thermal, which means that the answer to the question above is negative and the information about the initial state of black hole is lost forever since a thermal black hole  radiation does not carry any information about it. In this way, the outside observer can only describe the black hole by a density matrix and we have a case where matter can evolves from a pure state to a mixed state. However, it contradict our basic knowledge about quantum mechanics since, there, a pure state can only evolve into an other pure state because of the unitarity of evolution operator $U$, $U^{\dagger}U$ = $1$. This is the black hole information loss paradox. 

Let us list some possibilities to solve this paradox, with its respectives problems:

$i)$ The evolution is indeed non-unitary and we must revise the basics concepts of quantum mechanics \cite{sw.hawking-prd14, sw.hawking-cmp87}. The main difficulties of this proposal seems to be the violation of energy conservation and the absence of an empty vacuum as ground state \cite{dj.gross-npb236, jr.ellis-npb241, t.banks-npb244, a.strominger-hepth9410187}.

$ii)$ The radiation is not thermal and carries information. Here we need new physics which is radically different from the one we know: we must leave concepts as locality and causality, since matter behind the horizon has to influence matter outside the horizon \cite{dn.page-prl44, g.thooft-npb256, sb.giddings-mpla22}.

$iii)$ information is stored in a stable black hole remnant \cite{y.aharonov-plb191}.
$iv)$ information is stored in a topological disconnected region which borns inside of the black hole in a topological change process \cite{sdh.hsu-plb644}. This process is also non-unitary and therefore suffers of the same problems of the first proposal. The other objection against topology change is the violation of cluster deposition(locality) \cite{l.susskind-hepth9405103,sdh.hsu-plb644}.

In this paper we shall focus in the topology change process to solve the black hole information paradox. We argue that using the fuzzy sphere model the objections against topology change above can be solved in a precise way.

\section{The fuzzy sphere model}
The fuzzy sphere $S_{F}^{2}$ was introduced in \cite{j.madore-cqg9} and has been treated extensively in the literature \cite{j.madore-aincdgpa, bp.dolan-jhep0502, ap.balachandran-ijmpa19, j.madore-cqg9}.   To obtain it  we must replace the commutative algebra of functions on a sphere $S^{2}$ by the  noncommutative algebra of matrices which is obtained by quantizing the coordinates $x_{\m}(i = 1,2,3)$  performing the transformation:
$x^{\m} \rightarrow  \hat{x}^{\m} = k\hat{J}^{\m}$ , where $\hat{J}^{\m}$ form the n-dimensional irreducible representation of the algebra of $SU(2)$ and
\begin{equation}
k = \frac{r}{\sqrt{n^{2} - 1}}.
\end{equation}
Therefore, the coordinates of the fuzzy sphere satisfies the commutation relation:
\begin{equation}
[\hat{x}^{\m},\hat{x}^{\n}] = i \lambda \!\!\! \slash C^{\m\n\alpha}\hat{x}^{\alpha} ,
\end{equation}
where $C^{\m\n\alpha} = r^{-1} \varepsilon^{\m\n\alpha}$ and $\lambda \!\!\! \slash$, which has a dimension of $(length)^{2}$, plays here a role analogous to that played by Planck's constant in quantum mechanics. The commutative limite is given by $\lambda \!\!\! \slash$ $\rightarrow$ $0$.

Then we have that the $S_{F}^{2}$ algebra is $Mat(2J + 1)$ and depends on the irreducible representation $J$ of the $SU(2)$ algebra and the operators $\hat{x}^{\m}$ act on the $(2J + 1)$-dimensional Hilbert space $H$ which is $Mat(2J + 1)$, where the scalar product is given by:
\begin{equation}
< M,N > = tr(M^{\dagger}N).
\end{equation}

One can show that $S_{F}^{2}$ posses a Hopf algebra structure \cite{ap.balachandran-ijmpa19} in a way that we can define a linear operation(the coproduct of Hopf algebra) on $S_{F}^{2}$ and compose two fuzzy spheres preserving algebraic proprieties intact. This coproduct of Hopf algebra, which we shall represent by $\Delta$, produces a topology change process where a fuzzy sphere splits in other two fuzzy spheres \cite{ap.balachandran-ijmpa19}.

Let M describes a wave function on $S_{F}^{2}$, the coproduct $\Delta: S_{F}^{2}(J) \rightarrow S_{F}^{2}(K) \otimes S_{F}^{2}(L)$ acts on M as \cite{ap.balachandran-ijmpa19}

\begin{eqnarray}
\Delta_{(K,L)} (M) = \sum_{\!\!\!\!\! \m_{1}, \m_{2},m_{1},m_{2}}C(K,L,J;\m_{1}, \m_{2})& \hspace{-37mm} C(K,L,J; m_{1}, m_{2}) \label{d-split}  \\ \nonumber &\times M_{\m_{1} + \m_{2}, m_{1} + m_{2}}e^{\m_{1}m_{1}}(K) \otimes e^{\m_{2}m_{2}}(L) \label{b-process}
\end{eqnarray}
and M $\in$ $S_{F}^{2}(J)$ splits into a superposition of wavefunctions on $S_{F}^{2}(K) \otimes S_{F}^{2}(L)$  in a way that the information in M is divided between the two fuzzy spheres with spins $K$ and $L$ respectivally.

We have that the process \equ{d-split} posses the following proprieties:

\begin{equation}
\Delta_{(K,L)}(M^{\dagger}) = \Delta(M)^{\dagger} , \label{p-1}
\end{equation}

\begin{equation}
\Delta_{(K,L)}(MN) = \Delta_{(K,L)}(M)\Delta_{(K,L)}(N) , \label{p-2}
\end{equation}

\begin{equation}
Tr\Delta_{(K,L)}(M) = TrM , \label{p-3}
\end{equation}

and

\begin{equation}
<\Delta_{(K,L)}(M), \Delta_{(K,L)}(N)> = <M, N>. \label{p-4}
\end{equation}
The two last proprieties assure that \equ{d-split} is a unitary. 
From equations \equ{p-1} and \equ{p-2} we have that if $A$ is a self-adjoint operator and $M$  is a wave function on $S_{F}^{2}(J)$, with $A$ acting on it, after the branching process, we must have $\Delta_{(K,L)}(A)$ and $\Delta_{(K,L)}(M)$. The operator $\Delta_{(K,L)}(A)$ is self-adjoint and act on $\Delta_{(K,L)}(M)$ as
\begin{equation}
\Delta_{(K,L)}(A)\Delta_{(K,L)}(M) = \Delta_{(K,L)}(AM). 
\end{equation}
Then if $M$ have a definite eigenvalue for $A$ on $S_{F}^{2}(J)$, then $\Delta_{(K,L)}(M)$ is a wave function with the same eigenvalue for $\Delta_{(K,L)}(A)$. It means that every operator on $S_{F}^{2}(J)$ is a constant of motion for the branching process and since the coproduct $\Delta$ is linear, if one quantity is conserved in $S_{F}^{2}(J)$ it is conserved after the branching process yet. 

One important consequence of this is that the braching process preserves cluster decomposition theorem which guarantees locality in physics. According to this theorem, the vacuum expectation value of a product of many operators - each of them being either in different regions A and B, where A and B are very separated - asymptotically equals the product of the expectation value of the product of the operators in A, times a similar factor from the region B. Consequently, sufficiently separated regions behave independently.
If $A_{1}, ..., A_{n}$ are n operators each localized in a bounded region and we pick some subset of the n operators to translate $\textbf{x}_{i}$ into $\textbf{x}^{'}_{i}$ = $\textbf{x}_{i}$ + $\rho \textbf{a}$,

\begin{eqnarray}
\lim_{\rho \rightarrow \infty}\hspace{-10mm}&  <M_{0},A_{1}(x_{1}),A_{2}(x_{2}),...,A_{j-1}(x_{j-1}),A_{j}(x^{'}_{j}) ,...A_{n}(x^{'}_{n})M_{0}> = \\ \nonumber   &<M_{0},A_{1}(x_{1}),A_{1}(x_{1}),...,A_{j-1}(x_{j-1})M_{0}> \times <M_{0},A_{j}(x^{'}_{j}),...A_{n}(x^{'}_{n})M_{0}>, \label{c-d}
\end{eqnarray}\\
where $M_{0}$ represents the vacuum state.
 
Let us suppose that the equation above is valid on $S_{F}^{2}(J)$, then on $S_{F}^{2}(K) \otimes S_{F}^{2}(L)$ we have 

\begin{eqnarray*}
\lim_{\rho \rightarrow \infty} \hspace{-4mm} &<\Delta(M_{0}),\Delta(A_{1}(x_{1})),...,\Delta(A_{j-1}(x_{j-1})),\Delta(A_{j}(x^{'}_{j}),...\Delta(A_{n}(x^{'}_{n}))\Delta(M_{0})> 
\end{eqnarray*} \vspace{-6mm}
\begin{eqnarray*}
= \lim_{\rho \rightarrow \infty} \hspace{-4mm} &<M_{0},A_{1}(x_{1}),A_{2}(x_{2}),...,A_{j-1}(x_{j-1}),A_{j}(x^{'}_{j}),...A_{n}(x^{'}_{n})M_{0}> \\ \nonumber 
\end{eqnarray*}  \vspace{-10mm}
\begin{eqnarray*}
&= <M_{0},A_{1}(x_{1}),A_{1}(x_{1}),...,A_{j-1}(x_{j-1})M_{0}>\times <M_{0},A_{j}(x^{'}_{j}),...A_{n}(x^{'}_{n})M_{0}> 
\end{eqnarray*} \vspace{-6mm}
\begin{center}
$= <\Delta(M_{0}),\Delta(A_{1}(x_{1})),...,\Delta(A_{j-1}(x_{j-1})\Delta(M_{0})>  \times$
\end{center} 
\begin{center}
$<\Delta(M_{0}),\Delta(A_{j}(x^{'}_{j})),...\Delta(A_{n}(x^{'}_{n}))\Delta(M_{0})>.$
\end{center}

In a way that the branching process \equ{b-process} preserves cluster decomposition and locality is not violated.

In this section we have arrived to a important result which is: we can define a topological change process, without break unitary or locality, where the information initially on a fuzzy sphere $S^{2}_{F}(J)$ is divided between two disconnected regions. This result will be essential in the next discutions.

\section{The black hole horizon as a fuzzy sphere}
Modelling the black hole horizon by a fuzzy sphere, let us suppose that M describes a wave function on $S_{F}^{2}$. Invoking the holographic principle \cite{l.susskind-jmp36}, where the physics inside of black hole can be described by a theory on the its horizon, we have that a full description of the initial state that collapsed into the black hole can be given by $M$.

In sections above we have shown that we can define, through the Hopf coproduct $\Delta$, a topological change process for the fuzzy sphere which conserves locality and unitarity. Modelling the black hole horizon by a fuzzy sphere we can use this process to address the information paradox. We have that in this process the information of the initial state, described by the wave function $M$ on a fuzzy sphere with spin $J$, and which collapsed into a black hole, is divided into two regions. One  of them is a fuzzy sphere with spin $K$, which we shall consider as the original world and name it ``the main word''. The other is a fuzzy sphere with spin $L$ which we shall name ``the baby world''. 

From the last section, the process above respects locality. The Hilbert space describing the entire universe is $H = H_{baby} \otimes H_{main}$ in which the wave function $M$ evolves unitarily. However an observer in the main words can not access the degrees of freedon of the other one, in a way that, for this observer his world appears to evolve from a pure to mixed state in a non unitary process and M cannot describe a wave function anymore but a density matrix . In this way the observer in the main world measure a horizon entropy given by
\begin{equation}
S = - k_{B}Tr[M \ln M] = k_{B} \ln(2K +1), \label{fs-entropy}
\end{equation}

Note that the formula \equ{fs-entropy}  gives us for the case of $k = 0$ (black hole does not have any degree of freedon), $S = 0$ which is naturally expected, unlike the one proposed in the reference \cite{bp.dolan-jhep0502}.

The area of the horizon is by the Bekenstein-Hawking formula \equ{B-H-form}
\begin{equation}
A_{K} = 4\frac{l_{p}}{k_{B}}^{2}S = 4l_{p}^{2}\ln(2K +1) \ ,
\end{equation}
and the mass spectrum, for a non-rotating black hole, is given by
\begin{equation}
M_{K}^{2} = \frac{A_{J}}{16 \pi} = \frac{1}{4 \pi}l_{p}^{2}\ln(2K +1). \label{bh-mass}
\end{equation}

We can choose the splitting process \equ{d-split} in a way that $K = J - \frac{1}{2}$, in the main world, and from \equ{bh-mass} this process will result in a decrease of mass of black hole in a logarithmic rate. Therefore, the splitting process \equ{d-split} can be used to describe the black hole evaporation process what is seen by the observer in the main world in a non-unitary way. In this way the information is lost to a topological disconnected region in the black hole evaporation process, and the Hawking radiation is thermal.
Note that M is, at all times, a wave function in $H$ and not a density matrix. A mixed state is only obtained if, in order to obtain a description of the system in terms of the degrees of freedon remaining in the local observer word, one traces over those degrees of freedon which fall past the horizon, for example, if we trace over $H_{baby}$  to obtain a density matrix valued only on $H_{main}$. In this way there is no violation of the unitarity.

\section{Conclusion}
In this paper we have modeled a black hole horizon by a fuzzy sphere and shown that we can get a topological change process which can be used to solve the black hole information paradox without break unitarity or locality. In this process a black hole event horizon, modeled by a fuzzy sphere with spin $J$, splits into two others. The information about the black hole initial state is divided between two topologically disconnected regions: the main and the baby world. An observer in the main word sees this process, which on his point of view, occur in a non-unitary way, due the impossibility of access the degrees of freedom of the baby world. However as we have shown the evolution of main and baby worlds together is unitary.

\section{Acknowledgements}

The author thanks to R.R. Landim for the carefull reading of the manuscript



\begin{thebibliography}{10}
\expandafter\ifx\csname url\endcsname\relax
  \def\url#1{\texttt{#1}}\fi
\expandafter\ifx\csname urlprefix\endcsname\relax\def\urlprefix{URL }\fi

\bibitem{sw.hawking-cmp43}
S.W.~Hawking, Particle Creation by Black Holes,
  Commun.Math.Phys. 43 (1975) 199--220.

\bibitem{sw.hawking-prd14}
S.W.~Hawking, Breakdown of Predictability in Gravitational Collapse,
  Phys.Rev.D 14 (1976) 2460--2473.

\bibitem{j.preskill-hepth9209058}
J.~Preskill, Do black holes destroy information?,
  hepth/9209058.

\bibitem{dn.page-hepth9305040}
D.N.~Page, Black hole information,
   hepth/9305040.

\bibitem{jd.bekenstein-lnc4}
J.D.~Bekenstein, Black holes and the second law,
  Lett. Nuovo Cim. 4 (1972) 737--740.


\bibitem{j.madore-aincdgpa}
J.~Madore, An introduction to non-commutative differential geometry and its physical applications, Cambridge University Press, Cambridge 1999

\bibitem{bp.dolan-jhep0502}
B.P.~Dolan,  Quantum black holes: The Event horizon as a fuzzy sphere,
  JHEP 0502:008 (2005).

\bibitem{h.grosse-ijtp35}
H.~Grosse, C.~Klimcik, P.~Presnajder,  Towards finite quantum field theory in noncommutative geometry,
  Int.J.Theor.Phys. 35 (1996) 231--244.

\bibitem{h.grosse-plb283}
H.~Grosse, J.~Madore, A Noncommutative version of the Schwinger model,
  Phys.Lett.B 283 (1992) 218--222.

\bibitem{h.grosse-lmp33}
H.~Grosse, P.~Presnajder, The Dirac operator on the fuzzy sphere,
  Lett.Math.Phys. 33 (1995) 171--182.

\bibitem{h.grosse-cmp185}
H.~Grosse, C.~Klimcik, P.~Presnajder, Field theory on a supersymmetric lattice,
  Commun.Math.Phys. 185 (1997) 155--175.

\bibitem{ap.balachandran-ijmpa19}
A.P.~Balachandran, Topology change for fuzzy physics: Fuzzy spaces as Hopf algebras,
  Int.J.Mod.Phys.A 19 (2004) 3395--3408.


\bibitem{sdh.hsu-plb644}
S.D.H.~Hsu, Spacetime topology change and black hole information,
  Phys.lett.B 664 (2007) 67--71.

\bibitem{sw.hawking-cmp87}
S.W.~Hawking , The Unpredictability of Quantum Gravity,
  Commun.Math.Phys. 87 (1982) 395. 

\bibitem{dj.gross-npb236}
D.J. Gross , IS QUANTUM GRAVITY UNPREDICTABLE?,
  Nucl.Phys.B 236 (1984) 349 

\bibitem{jr.ellis-npb241}
J.R.~Ellis, J.S.~Hagelin, D.V.~Nanopoulos, M.~Srednicki,  Search for Violations of Quantum Mechanics,
  Nucl.Phys.B 241 (1984) 381. 

\bibitem{t.banks-npb244}
T.~Banks, L.~Susskind, M.E.~Peskin,  Difficulties for the Evolution of Pure States Into Mixed States,
  Nucl.Phys.B 244 (1984) 125. 

\bibitem{a.strominger-hepth9410187}
A.~Strominger, Unitary rules for black hole evaporation,
  hep-th/9410187. 

\bibitem{dn.page-prl44}
D.N.~Page, IS BLACK HOLE EVAPORATION PREDICTABLE?,
  Phys.Rev.Lett. 44 (1980) 301. 

\bibitem{g.thooft-npb256}
G.'t Hooft , On the Quantum Structure of a Black Hole,
  Nucl.Phys.B 256 (1985) 727. 

\bibitem{sb.giddings-mpla22}
S.B.~Giddings, Black holes, information, and locality,
  Mod.Phys.Lett.A 22 (2007) 2949-2954. 

\bibitem{y.aharonov-plb191}
Y.~Aharonov, A.~Casher, S.~Nussinov, The Unitarity Puzzle And Planck Mass Stable Particles,
  Phys.Lett.B 191 (1987) 51.

\bibitem{l.susskind-hepth9405103}
L.~Susskind,  Comment on a proposal by Strominger,
  hep-th/9405103.

\bibitem{l.susskind-jmp36}
L.~Susskind, The World as a hologram,
  J.Math.Phys. 36 (1995) 6377-6396.

\bibitem{j.madore-cqg9}
J.~Madore, The Fuzzy sphere,
  Class.Quant.Grav. 9 (1992) 69--88.

\end{thebibliography}
\end{document}